\documentclass[preprints,article,accept,pdftex,moreauthors]{./Definitions/mdpi}

\usepackage{amsmath}
\usepackage{amssymb}
\usepackage{braket}
\usepackage[normalem]{ulem}

\firstpage{1} 
\pubvolume{1}
\issuenum{1}
\articlenumber{0}
\pubyear{2022}
\copyrightyear{2022}
\externaleditor{Academic Editor: Antonio M. Scarfone} 
\datereceived{17 January 2022} 
\dateaccepted{9 February 2022} 
\datepublished{} 
\hreflink{https://doi.org/} 

\Title{Zonal Estimators for Quasiperiodic Bosonic Many-Body Phases}

\TitleCitation{Zonal Estimators for Quasiperiodic Bosonic Many-Body Phases}

\Author{Matteo Ciardi 
 $^{1,2,}$ \orcidA{}, Tommaso Macr\`i $^{3,4}$\orcidB{} and Fabio Cinti $^{1,2,5}$\orcidC{}}

\AuthorNames{Matteo Ciardi, Tommaso Macr\`i and Fabio Cinti}

\AuthorCitation{Ciardi, M.; Macr\`i, T.; Cinti, F.}

\address{$^{1}$ \quad Dipartimento di Fisica e Astronomia, Universit\`a di Firenze, I-50019 Sesto Fiorentino, Italy\\
$^{2}$ \quad {INFN}, Sezione di Firenze, I-50019 Sesto Fiorentino, Italy\\
$^{3}$ \quad Departamento de F\'isica Te\'orica e Experimental and International Institute of Physics, Universidade Federal do Rio Grande do Norte, Natal 59078-970, RN, Brazil;\\
$^{4}$ \quad Harvard-Smithsonian Center for Astrophysics, Institute for Theoretical Atomic, Molecular and Optical Physics (ITAMP), Cambridge, MA 02138, USA\\
$^{5}$ \quad Department of Physics, University of Johannesburg, P.O. Box 524, Auckland Park 2006, South Africa}

\corres{Correspondence: matteo.ciardi@unifi.it} 

\abstract{In this work, we explore the relevant methodology for the investigation of interacting systems with contact interactions, 
and we introduce a class of zonal estimators for path-integral Monte Carlo methods, designed to provide physical information about limited regions of inhomogeneous systems.
We demonstrate the usefulness of zonal estimators by their application to a system of trapped bosons in a quasiperiodic potential in two dimensions, focusing on finite temperature properties across a wide range of values of the potential. Finally, we comment on the generalization of such estimators to local fluctuations of the particle numbers and to magnetic ordering in multi-component systems, spin systems, and systems with nonlocal interactions.}


\begin{document}

\section{Introduction}\label{sec:1}

Path-integral Monte Carlo (PIMC) methods~\cite{0034-4885-75-9-094501} are of great importance for the simulation of strongly correlated systems where other techniques fail, especially in two and three spatial dimensions. Over~the last thirty years, this has been amply demonstrated on quantum fluids~\cite{bps06,zen14,PhysRevB.84.094119} and, more recently, in~ultra-cold gases like, for~instance, dipolar systems~\cite{Buchler2007,PhysRevA.96.013627,PhysRevLett.113.240407,PhysRevLett.119.215302,https://doi.org/10.7566/JPSJ.85.053001} and Rydberg atoms~\cite{Cinti:2014aa,PhysRevLett.104.223002,Cinti2010b,1367-2630-16-3-033038}. For~strongly-interacting quantum fluids, there is at present considerable interest towards the exploration of patterns owing to peculiar symmetries such as quantum-cluster crystals~\cite{PhysRevA.87.061602,Macri:2014aa}, stripe phases~\cite{C7SM00254H}, or~cluster quasicrystals~\cite{Barkan2014,Barkan2011,PhysRevB.101.134522}, with~the aim of understanding fundamental physical phenomena. In~this regard, also thanks to the increase in computational capabilities, advancements in PIMC methods continue to play a key~role.

In this work, we detail the numerical techniques required to investigate the quantum properties of interacting trapped bosons in an external quasiperiodic potential at finite temperature, with~specific attention to superfluidity.
Quasicrystals are a fascinating state of solid-state matter exhibiting behaviors halfway between a periodically ordered structure and a fully disordered system. They were first synthesized in 1982 (and their discovery later announced in 1984) by Shechtman~et~al.~\cite{PhysRevLett.53.1951}. Later, Bindi~et~al. demonstrated that quasicrystals can also originate naturally, in~the presence of extreme conditions such as collisions between asteroids~\cite{Bindi2009}. Their properties have already been the subject of extensive theoretical investigation~\cite{Lev84}, motivated in part by the discovery of aperiodic tilings that can cover the plane without being bounded by the symmetries of classical crystallography, such as the Penrose tiling~\cite{Pen74}.~At~finite temperature, the~thermodynamic features of quasicrystals can be established in terms of the interplay between different length and energy scales pertaining to the inter-particle potentials~\cite{PhysRevE.70.021202}. These classical systems were found to remain stable even at zero temperature~\cite{Dotera2014}. 
Quasicrystalline properties have been observed in a variety of physical systems, for~instance, in~nonlinear optics~\cite{PhysRevE.66.046220,PhysRevLett.82.4627,PhysRevLett.74.258}, on~twisted bilayer graphene~\cite{Ahn2018} and in ultra-cold trapped atoms~\cite{Sbroscia2020,Viebahn2019}. In~the latter case, quasicrystalline structures generated by means of optical lattices are employed to experimentally investigate remarkable effects such as many-body localization in one and two-dimensions~\cite{Choi2016}, and~have been suggested as a candidate to probe the existence of two-dimensional Bose glasses~\cite{Fisher1989}. In~this regard, recent PIMC simulations support the existence of a Bose glass phase, fully stable and robust at finite temperature, in~a region of parameters suitable for experimental setups~\cite{cia22,soy11}. Other works have delineated zero-temperature phase diagrams, in~the mean-field approximation as well as for a strong interactions using ab-initio techniques~\cite{PhysRevA.72.053607,gau21,joh19,sza20}. 

Here, we summarize the derivation of the pair-product approximation for particles interacting through hard-core interactions in two or three dimensions, and~we present the details of our implementation. Then, we explore a new zonal estimator, which gives access to local information about the superfluidity in finite regions of trapped systems, and~is therefore well-suited to the study of spatially inhomogeneous potentials. 
Zonal estimators can be relevant to the detection of correlated phases, such as the Bose glass phase, which is characterized by rare regions where superfluidity and finite compressibility coexist~\cite{cia22}.

This paper is organized as follows. In~Section~\ref{sec:3}, we present and discuss the PIMC methodology for ensembles of interacting bosons through the pair-product approximation. In~Section~\ref{sec:2}, we introduce a model Hamiltonian describing interacting trapped bosons subjected to a quasiperiodic potential. Structural properties such as density profiles and diffraction patterns are shown in Section~\ref{sec:4}, whereas we examine global quantum features in Section~\ref{sec:5}. The~zonal estimator of the superfluid fraction is explored in Section~\ref{sec:5a}. To~conclude, Section~\ref{sec:6} is devoted to the discussion of our findings, drawing some~conclusions. 

\section{Methodology}\label{sec:3}

In this section, we review the implementation of the PIMC to the study of an interacting Bose gas in an external potential at finite temperature. 
This methodology aims to sample the partition function of a quantum system at finite temperature. 
In line with Feynman's path integral theory~\cite{fey98, fey10}, thermodynamic properties are addressed by considering an equivalent classical system, in~which each quantum particle is represented by a classical polymer. As~a result, quantum quantities, like for instance superfluidity or Bose--Einstein condensation, can be mapped across the equivalence as properties of the polymers themselves~\cite{krauth2006statistical}. The~evaluation of those quantities takes place via a standard classical Monte Carlo procedure such as the Metropolis algorithm~\cite{9780511614460}, allowing us to sample thermodynamic properties within a precision limited only by numerical and statistical errors.
At present, one of the most efficient ways of sampling configurations of connected polymers is operated through the so-called worm algorithm~\cite{bps06,PhysRevLett.96.070601}; originally developed for the grand-canonical ensemble, we routinely use the worm algorithm in its canonical version to sample superfluid fraction, condensate fraction, or~ground state~energy. 

In the following, we recall the formalism and derivations at the core of PIMC. The~partition function, ${\cal Z}$, is defined as the trace of the equilibrium density matrix operator, $\rho$, at~ temperature $T$ Ref.~\cite{Ceperley1995}:
\begin{equation}\label{eq:partf1}
\rho = e^{-\beta\, \hat{{\cal H}}} \, , \qquad {\cal Z}=\text{Tr~}e^{-\beta\, \hat{{\cal H}}} ,
\end{equation}
$\beta$ being the inverse temperature parameter, $\beta=1/k_{\text{B}}T$.

For $N$ distinguishable particles, denoting with $\textbf{r}_i$ the position of the $i$-th particle, and~introducing $\textbf{R}$ = $(\textbf{r}_{1},\textbf{r}_{2},\ldots,\textbf{r}_{N})$, we can project the density matrix operator on the basis of spatial coordinates $\ket{\textbf{R}}$, obtaining
\begin{equation}\label{eq:partf2}
\rho(\textbf{R},\textbf{R},\beta) = \braket{\textbf{R} | e^{-\beta\, \hat{{\cal H}}} | \textbf{R}} \, , \qquad {\cal Z} = \int d\textbf{R}\, \rho(\textbf{R},\textbf{R},\beta).
\end{equation}

For an ensemble of bosons, taking into account permutations, we arrive at
\begin{equation}\label{eq:partf3}
{\cal Z} = \frac{1}{N!} \sum_{P} \int d\textbf{R}\, \rho(\textbf{R},P\textbf{R},\beta),
\end{equation}
where $P\textbf{R}$ = $(\textbf{r}_{P(1)},\textbf{r}_{P(2)},\ldots,\textbf{r}_{P(N)})$ denotes a permutation of the particle~coordinates.

Introducing a decomposition of the density matrix operator into a convolution of density matrices at a higher temperature, Equation~\eqref{eq:partf4} yields
\begin{equation}\label{eq:partf4}
{\cal Z}=\frac{1}{N!}\sum_{P}\int d\textbf{R}^0d\textbf{R}^1\ldots d\textbf{R}^{M-1} \; \rho(\textbf{R}^0,\textbf{R}^1,\tau) \rho(\textbf{R}^1,\textbf{R}^2,\tau) \cdots\rho(\textbf{R}^{M-1},P\textbf{R}^0,\tau),
\end{equation}
with $\beta$ breaking up into $M$ smaller intervals $\tau = \beta /M$, and~ $\textbf{R}^m = (\textbf{r}_{1}^{m},\textbf{r}_{2}^{m},\ldots,\textbf{r}_{N}^{m})$ the coordinates of particles on a given time slice.  To~each particle $\textbf{r}_i$ corresponds, then, a~classical polymer made of $m=1,2,\dots, M$ beads, connected with each other through harmonic springs~\cite{Ceperley1995}. Errors introduced by the equivalence are reduced as $M$ increases. Likewise, the~same decomposition can be applied to the evaluation of observables by Monte Carlo~sampling. 

For a generic diagonal observable, $\hat{\cal A}$ such that $\braket{\textbf{R}|\hat{\cal A}|\textbf{R}'} = \cal A(\textbf{R}) \delta(\textbf{R}-\textbf{R}')$, it follows~that
\begingroup\makeatletter\def\f@size{9}\check@mathfonts
\def\maketag@@@#1{\hbox{\m@th\normalsize\normalfont#1}}%

\begin{equation}\label{eq:obs}
\langle {\cal A} \rangle = \frac{1}{{\cal Z}N!}\sum_{P} \int d\textbf{R}^0d\textbf{R}^1\ldots d\textbf{R}^{M-1} \; {\cal A}(\textbf{R}^0)\, \rho(\textbf{R}^0,\textbf{R}^1,\tau)\rho(\textbf{R}^1,\textbf{R}^2,\tau)\cdots\rho(\textbf{R}^{M-1},P\textbf{R}^0,\tau).
\end{equation}
\endgroup

Equation~\eqref{eq:obs} is evaluated through a stochastic process, consisting of the generation of random configurations $\{\textbf{R}^0, \textbf{R}^1, \dots \textbf{R}^{M-1}\}$ from the probability distribution
\begin{equation}\label{eq:distribution}
\pi(\textbf{R}^0, \textbf{R}^1, \dots \textbf{R}^{M-1}) = \frac{1}{{\cal Z}N!}\sum_{P}  \rho(\textbf{R}^0,\textbf{R}^1,\tau)\rho(\textbf{R}^1,\textbf{R}^2,\tau)\cdots\rho(\textbf{R}^{M-1},P\textbf{R}^0,\tau).
\end{equation}

The thermodynamic average then is measured as an average of $\{{\cal A}(\textbf{R}^0) \}$ over the sampled configurations~\cite{Boninsegni2005}. 

Having to employ a finite number of time slices, $M$, the~most sensitive step of the procedure lies in finding a good approximation of the high-temperature density-matrix elements $\rho(\textbf{R}^m,\textbf{R}^{m+1},\tau)$ in \eqref{eq:partf4} \cite{Chin1997,Boninsegni2005,pil06}. Due to the nature of the two-body interaction potential between the bosons in Hamiltonian $V_{int}(|\hat{\textbf{r}}_i-\hat{\textbf{r}}_j|)$ (see~\eqref{eq:ham} for an application), and~to the density regime of interest, in~the present work we apply a pair-product approximation (PPA) ansatz~\cite{Ceperley1995}. 
The rest of this section is devoted to the treatment and implementation of contact interactions in this context; similar derivations and other details can be found, e.g.,~in~\cite{pil06, pil08} and references therein, and~the supplemental material of~\cite{gau21}.

We express the density-matrix terms as
\begin{equation}\label{eq:rho1}
 \rho(\textbf{R}, \textbf{R}^\prime; \tau) = \rho_{free} (\textbf{R}, \textbf{R}^\prime; \tau) \prod_{i<j} \frac{\rho_{int}^{rel}(\textbf{r}_{ij},\textbf{r}{^\prime}_{ij};\tau)}{\rho_{free}^{rel}(\textbf{r}_{ij},\textbf{r}^\prime_{ij};\tau)} + O(\tau^2)\,,
 \end{equation} 
(to keep Equation~\eqref{eq:rho1} simple, we omit the indices $m$). Here,
\begin{equation}
    \rho_{free} (\textbf{R}, \textbf{R}^\prime; \tau) = \braket{\textbf{R} | \exp\{ -\tau \hat{\mathcal{H}}_{free} \} | \textbf{R}'} = \prod_{i=1}^{N} \frac{1}{(4\pi\lambda_i \tau)^{d/2}} \exp \left\{-\frac{(\textbf{r} -\textbf{r}')^2}{4\lambda_i \tau} \right\} \,
\end{equation}
with $\lambda_i = \hbar^2 / 2m_i$, is the density matrix of the non-interacting Hamiltonian of $N$ particles, 
$ \hat{\mathcal{H}}_{free} = \sum_{i=1}^{N} \hat{\textbf{p}}_i^2 / 2m_i \,. $

For two particles, labeled $i$ and $j$, we can decompose the Hamiltonian into a center-of-mass term and a relative term, with~the relative term being $ \hat{\mathcal{H}}_{int}^{rel}  = \frac{\hat{\textbf{p}}_{ij}^2}{2m_r} + V_{int}(\hat{\textbf{r}}_{ij}) $; for free particles, the~relative Hamiltonian is only $\hat{\mathcal{H}}_{free}^{rel}  = \frac{\hat{\textbf{p}}_{ij}^2}{2m_r} $. Here we have introduced the relative coordinates $ \textbf{r}_{ij} = \textbf{r}_{j} - \textbf{r}_{i} $, $	\textbf{p}_{ij} = (m_i \textbf{p}_{i} - m_j \textbf{p}_{j} / (m_i + m_j) $,
and the reduced mass $m_r = m_i m_j / (m_i + m_j)$.
We can then write the propagators
\begin{align} \label{eq:prop_rel}
    \rho_{free}^{rel}(\textbf{r}_{ij},\textbf{r}{^\prime}_{ij};\tau) & = 
    \braket{\textbf{R} | \exp\{  -\tau \hat{\mathcal{H}}_{free}^{rel} \} | \textbf{R}'} = 
    \frac{1}{(4\pi\lambda_{r}\tau)^{d/2}}  \exp \left\{-\frac{|\textbf{r}_{ij} -\textbf{r}_{ij}'|^2}{4\lambda_{r}\tau} \right\}    \, , \\
    \rho_{int}^{rel}(\textbf{r}_{ij},\textbf{r}{^\prime}_{ij};\tau) & = 
    \braket{\textbf{R} | \exp\{  -\tau \hat{\mathcal{H}}_{int}^{rel} \}  | \textbf{R}'} \, ,
\end{align}
where $\lambda_{r} = \hbar^2 / 2 m_{r} = \lambda_i + \lambda_j $.

We use a standard Metropolis procedure, which consists of generating new configurations according to the free particle distribution, and~then accepting or rejecting them according to a statistical weight, which takes external potentials and interactions into account. The~form \eqref{eq:rho1} is best suited for this procedure, as~long as we can efficiently determine the terms under the product symbol. 
For ease of notation, we now take $\textbf{r}=\textbf{r}_{ij}$ and define
\begin{equation} 
\label{eq:rhoint}
\rho_{int}(\textbf{r}, \textbf{r}', \tau) = \frac{\rho^{rel}_{int}(\textbf{r}, \textbf{r}', \tau)}{\rho^{rel}_{free}(\textbf{r}, \textbf{r}', \tau)}\,.
\end{equation} 

In order to estimate $\rho^{rel}_{int}$ for the model proposed in Equation~\eqref{eq:ham}, we can expand it on the eigenfunctions of the relative Schr\"{o}dinger equation
\begin{equation}
\hat{\mathcal{H}}_{int}^{rel} \psi(\textbf{r}) = \left[ -\lambda_r \nabla^2_{\textbf{r}} + V_{int}(\textbf{r}) \right] \psi(\textbf{r}) = E \psi(\textbf{r}). 
\end{equation} 

For central potentials, which only depend on $r = |\textbf{r}|$, like the one considered in this study, the~equation splits into an angular part, giving rise to spherical harmonics in $d$ dimensions, and~a radial part:
\begin{equation}
\label{eq:schroed2d}
- \lambda_r \left[ \frac{\partial^2}{\partial r^2} + \frac{1}{r} \frac{\partial}{\partial r} - \frac{m^2}{r^2} \right]  u_{km}(r) + V_{int}(r) u_{km}(r) = \lambda_r k^2 u_{km}(r)\,, \qquad  d=2 
\end{equation}
\begin{equation} 
\label{eq:schroed3d}
- \lambda_r \left[ \frac{1}{r^2} \frac{\partial}{\partial r} \left( r^2 \frac{\partial}{\partial r} \right) - \frac{l(l+1)}{r^2} \right]  u_{kl}(r) + V_{int}(r) u_{kl}(r) = \lambda_r k^2 u_{kl}(r)\,. \qquad  d=3 
\end{equation}

In terms of these wavefunctions, we can expand the relative density matrix into
\begin{adjustwidth}{-0.2\extralength}{0cm}
\begin{multline}\label{inte2d}
\rho_{int}^{rel}(\textbf{r}, \textbf{r}', \tau)=\sum_{m=0}^{\infty} c_m \cos(m\theta) \int_0^\infty dk \; e^{-\lambda_r\tau k^2} u_{km}(r)^* u_{km}(r') +  \sum_n e^{-\tau E_n} \phi_n(\textbf{r})^* \phi_n(\textbf{r}')\,, \\ d=2 
\end{multline}
\begin{multline} \label{inte3d}
\rho_{ int}^{rel}(\textbf{r}, \textbf{r}', \tau)=\frac{1}{2\pi^2} \sum_{l=0}^{\infty} (2l+1)P_l(\cos \theta) \int_0^\infty dk \; k^2 e^{-\lambda_r\tau k^2} u_{kl}(r)^* u_{kl}(r')+ \sum_n e^{-\tau E_n} \phi_n(\textbf{r})^* \phi_n(\textbf{r}')\,.  \\ d=3 
\end{multline}
\end{adjustwidth}
The coefficients $c_m$ are defined as $c_0=1$, $c_m=2$ for $m>0$. The~functions $P_l(\cos \theta)$ are the Legendre polynomials of degree $l$ \cite{asmar}. Finally, the~$\phi_n(\textbf{r})$ are the bound states of the potential, if~any, and~$E_n$ their energies; they play no role in the study of repulsive potentials, such as the one we are~considering.

The free problem has straightforward solutions that, in~the two-dimensional case,~yield
\begin{eqnarray}
&&u_{km}(r) = \sqrt{\frac{k}{2\pi}} J_m(kr) \\
&&\rho_{free}^{rel}(\textbf{r}, \textbf{r}', \tau) = \sum_{m=0}^{\infty} c_m \cos(m\theta) \int_0^\infty dk \; e^{-\lambda_r\tau k^2} \frac{k}{2\pi} J_m(kr) J_m(kr')\,.
\end{eqnarray}

In three dimensions, we get
\begin{eqnarray}
&&u_{kl}(r) = j_l(kr) \\
&&\rho_{free}^{rel}(\textbf{r}, \textbf{r}', \tau) = \frac{1}{2\pi^2} \sum_{l=0}^{\infty} (2l+1)P_l(\cos \theta) \int_0^\infty dk \; k^2 e^{-\lambda_r\tau k^2} j_l(kr) j_l(kr')\,.
\end{eqnarray}

$J_m(x)$ are the Bessel functions of the first kind, and~$j_l(x) = \sqrt{\pi / 2x} J_{l+1/2}(x)$ are the spherical Bessel functions of the first kind; the factor appearing in the two-dimensional case is due to the normalization and orthogonality relations. In~both cases, the~sum can be computed analytically by employing tabulated integrals, leading back to the simple form of \eqref{eq:prop_rel}. 

More generally, it is necessary to solve Equations \eqref{eq:schroed2d} or \eqref{eq:schroed3d} numerically to find the eigenfunctions. If~the interaction is a short-range potential, so that $V_{int}(r)=0$ when $r > r_0$, for~some value of $r_0$, the~eigenfunctions in the region $r > r_0$ are a generalization of the free~case:
\begin{eqnarray}
u_{km}(r) &=&  \sqrt{\frac{k}{2\pi}} \left[ cos(\delta_m(k)) J_m(kr) - sin(\delta_m(k)) Y_m(kr) \right]\,,\qquad   d=2 \label{pippo}\\
\nonumber\\
u_{kl}(r) &=& cos(\delta_l(k)) j_l(kr) - sin(\delta_l(k)) y_l(kr)\,. \qquad \qquad \qquad \,\,\,\,\, d=3\label{pippo1}
\end{eqnarray}
$Y_m(x)$ and $y_l(x) = \sqrt{\pi / 2x} Y_{l+1/2}(x)$ are, respectively, the~Bessel and spherical Bessel functions of the second kind. For~$r < r_0$, it is still necessary to solve the Schrödinger equation. The~phase shifts $\delta_l$ are determined by imposing smoothness conditions on the wavefunction at $r=r_0$. In~the particular case of a hard-core potential of radius $r_0$, the~requirement is
\begin{eqnarray}
\tan \delta_m(k) &=& \frac{J_m(kr_0)}{Y_m(kr_0)}, \qquad  d=2\,\\
\tan \delta_l(k) &=& \frac{j_l(kr_0)}{y_l(kr_0)}. \qquad \,\, d=3\, 
\end{eqnarray}
In order to implement the above formalism efficiently in our simulations, we recast Equation~\eqref{eq:rhoint} as follows:
\begin{equation}
\begin{split}
\rho_{int}(\textbf{r}, \textbf{r}', \tau) &= \frac{\rho^{rel}_{free}(\textbf{r}, \textbf{r}', \tau) + \rho^{rel}_{int}(\textbf{r}, \textbf{r}', \tau) - \rho^{rel}_{free}(\textbf{r}, \textbf{r}', \tau)}{\rho^{rel}_{free}(\textbf{r}, \textbf{r}', \tau)} \\
&= 1 + \frac{\rho^{rel}_{int}(\textbf{r}, \textbf{r}', \tau) - \rho^{rel}_{free}(\textbf{r}, \textbf{r}', \tau)}{\rho^{rel}_{free}(\textbf{r}, \textbf{r}', \tau)}\\
&= \begin{cases}
    1 + \frac{1}{\rho^{rel}_{free}(\textbf{r}, \textbf{r}', \tau)} \sum_{m=0}^{\infty} c_m \cos(m\theta)\mathcal{I}_m(r,r')\,, & d=2  \\
    & \\
    1 + \frac{1}{\rho^{rel}_{free}(\textbf{r}, \textbf{r}', \tau)} \sum_{l=0}^{\infty} (2l+1)P_l(\cos \theta) \mathcal{I}_l(r,r')\,, & d=3\, 
     \,\label{eq:rho_expansion_2d}
  \end{cases}
\end{split}
\end{equation}
where
\begin{equation}\label{eq:integral_2d}
\mathcal{I}_m(r,r') =  \frac{1}{2\pi^2} \int dk \; e^{-\tau \lambda_r k^2} \left( u_{km}(r)^* u_{km}(r') - \frac{k}{2\pi} J_{m}(kr) J_{m}(kr') \right)\,, \qquad  \qquad d=2
\end{equation}
\begin{equation}\label{eq:integral_3d}
\mathcal{I}_l(r,r') = \frac{1}{2\pi^2} \int dk \; e^{-\tau \lambda_r k^2} k^2  \left( u_{kl}(r)^* u_{kl}(r') - j_{l}(kr) j_{l}(kr') \right)\,.  \qquad \qquad \qquad \,\,\,\,\,  d=3
\end{equation} 

The interacting propagator cannot be calculated analytically; the integrals must be computed numerically and tabulated before running the simulations. While, in~principle, we could tabulate the entire propagator as a function of $r$, $r'$, and~$\theta$, the~decomposition of \eqref{eq:rho_expansion_2d} has several advantages. First of all, it cleanly separates the contributions from the free propagator, which are relevant at any $m$ at large enough distances, so that we only need to write tables for those values of $m$ that actually present a variation with respect to the free case. Moreover, since the angular variable $\theta$ is explicitly considered in the sum, the~integrals only need to be tabulated as a function of $r$ and $r'$, reducing computational time and memory usage~considerably.

In our two-dimensional simulations, at~all temperatures, we have found the contributions from the harmonics $m\ge 1$ to be negligible, so that we only use
\begin{equation} \label{eq:propagator_l0}
\rho_{int}(\textbf{r}, \textbf{r}', \tau) =
1 + \frac{\mathcal{I}_0(r,r')}{\rho^{rel}_{free}(\textbf{r}, \textbf{r}', \tau)}\,.
\end{equation}

Having established the form of the propagator, we can now use it to compute values of thermodynamic observables. Some special care must be devoted to the thermal estimator of the total or kinetic energy, for~which the effective potential leads to a contribution of the~form
\begin{equation} 
\frac{\partial }{\partial \tau} u (\textbf{r}, \textbf{r}', \tau), 
\end{equation} 
with
\begin{equation} 
u (\textbf{r}, \textbf{r}', \tau) = -\ln \rho_{int}(\textbf{r}, \textbf{r}', \tau).
\end{equation}

For a generic $\rho_{int}$ of the form
\begin{equation} 
\rho_{int}(\textbf{r}, \textbf{r}', \tau) =
1 + \frac{\mathcal{I}(r,r')}{\rho^{rel}_{free}(\textbf{r}, \textbf{r}', \tau)}\,,
\end{equation}
with
\begin{equation} 
\mathcal{I} = \int dk \; e^{-\tau \lambda_r k^2} F(r,r',k),
\end{equation}
we can introduce
\begin{equation} 
\mathcal{J} = \lambda_r \int dk \; k^2\;  e^{-\tau \lambda_r k^2} F(r,r',k) ;
\end{equation} 
it is then possible to show that
\begin{equation} 
\frac{\partial }{\partial \tau} u (\textbf{r}, \textbf{r}', \tau) = \frac{1}{\rho^{rel}_{free}(\textbf{r}, \textbf{r}', \tau) + \mathcal{I}(r,r')} \left[ \mathcal{J}(r,r') + \frac{1}{\tau} \left[\frac{(\textbf{r} -\textbf{r}')^2}{4\lambda_r\tau} - \frac{d}{2} \right] \mathcal{I}(r,r') \right] , 
\end{equation} 
$d$ being the dimensionality of the system. In~particular, this applies to the propagator \eqref{eq:propagator_l0} used in the present~work.

\begin{figure}[t]
\includegraphics[width=10cm]{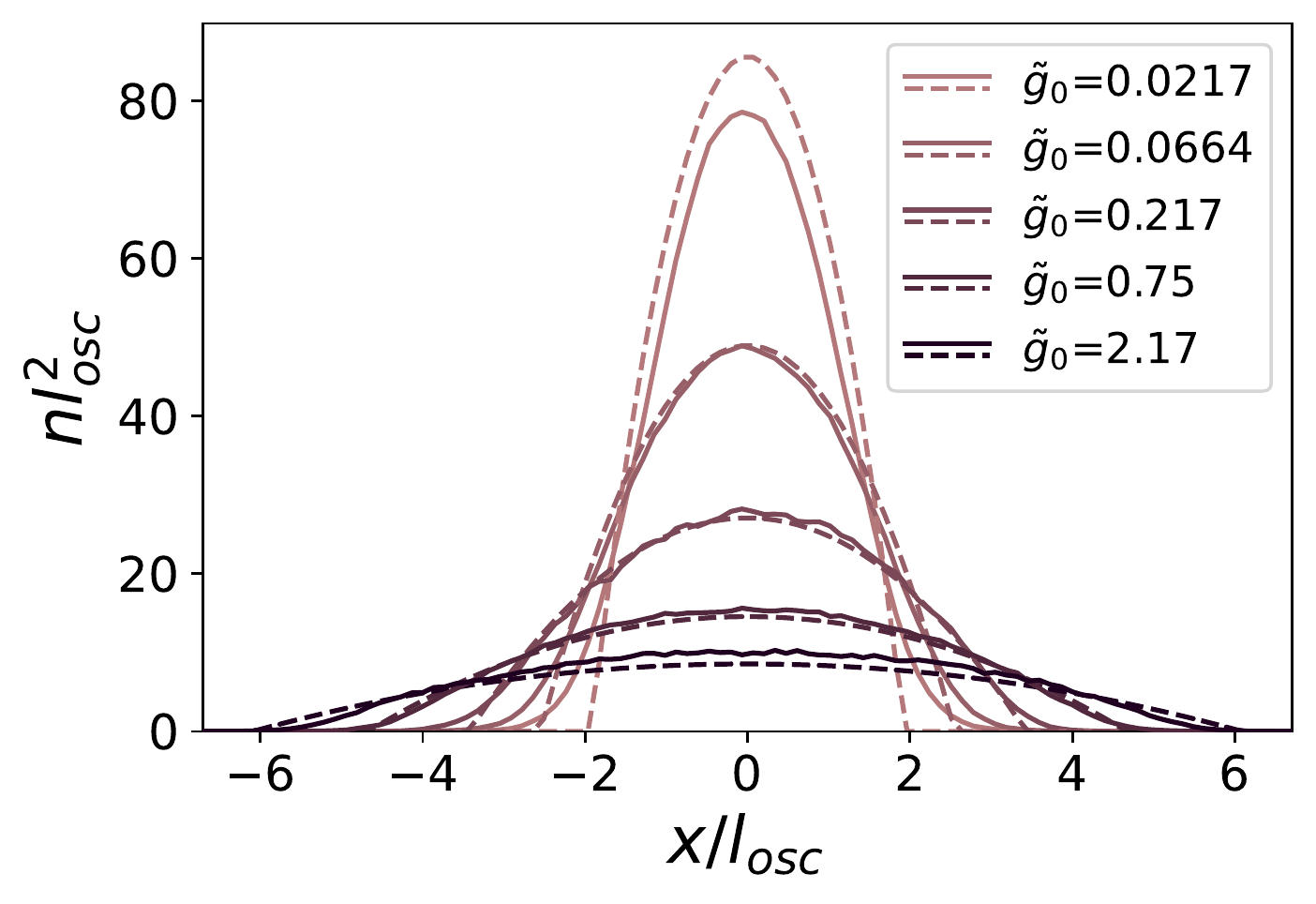}
\caption{Comparison between exact density profiles from PIMC simulations (solid lines) and those obtained from the 2D Thomas--Fermi approximation (dashed lines). The~shades of purple correspond to different values of the interaction, from~small (light) to strong (dark), as~indicated in the legend.
\label{fig0}}
\end{figure}   

As a final technical consideration, we note that the above treatment of the interaction has been carried out in the absence of external potentials, which represent, instead, a~crucial component of our physical problem. The~Trotter decomposition of \eqref{eq:partf4} allows us to treat different potential and interaction terms independently, or~to group them together as needed, as~long as we pick a fittingly small value of $\tau$. For~a problem in the harmonic trap, it is most efficient to sample configurations from the harmonic propagator
\begin{equation}
\rho_{osc}(\textbf{r},\textbf{r}',\tau) = \braket{ \textbf{r} | \exp \left\{ -\tau \frac{\hat{\textbf{p}}^2}{2m} + \frac{m \omega^2 \hat{\textbf{r}}^2}{2} \right\} | \textbf{r}' } \, ,  
\end{equation}
which can be computed analytically~\cite{krauth2006statistical, Krauth1996}. We obtained satisfying results by employing this distribution together with the pair-product propagator in the absence of external potentials \eqref{eq:rhoint}, as~displayed in Figure
~\ref{fig0}, so that the complete form of our propagator~is
\begin{equation} \label{eq:final_rho}
\rho(\textbf{R}, \textbf{R}^\prime; \tau) = \prod_{i=1}^{N} \rho_{osc} (\textbf{r}_i, \textbf{r}_i \prime; \tau) \prod_{i=1}^{N} e^{-\tau V_{ext}(\textbf{r}_i)} 
\prod_{i<j} \rho_{int}(\textbf{r}_{ij},\textbf{r}{^\prime}_{ij};\tau)
\end{equation}

\section{Application: Trapped Bosons in a Quasicrystal~Potential}\label{sec:2}

\begin{figure}[t]
\includegraphics[width=10.cm]{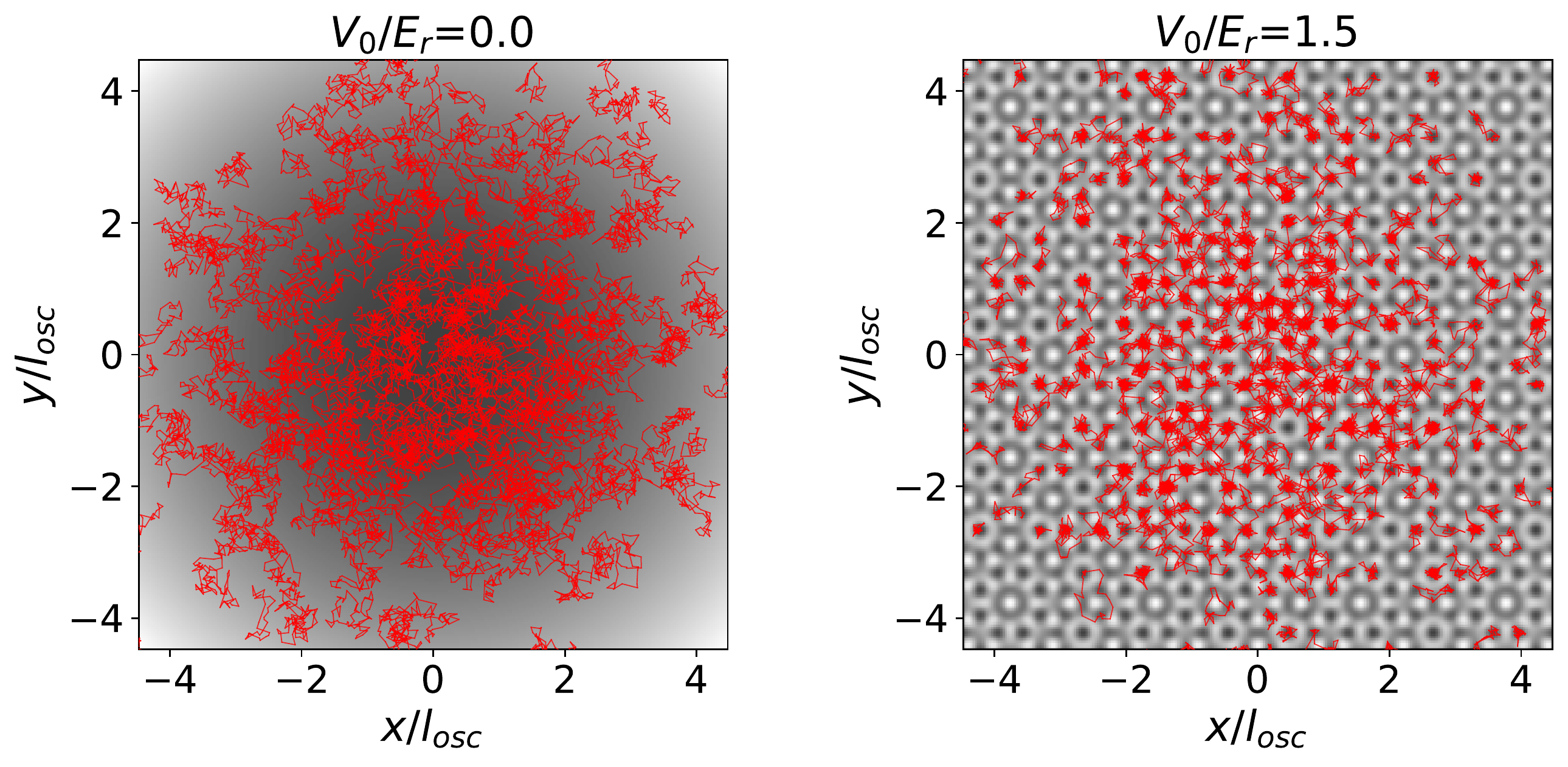}
\caption{Snapshots of particle configurations at $\tilde{g}_0=2.1704$, for~two values of $V_0$. Red lines represent the links between successive beads in the equivalent polymer system, as~described in Section~\ref{sec:3}. The~shaded background represents the potential, with~darker to brighter areas corresponding to lower to higher values. On~the \textbf{left}, we show a configuration in the absence of the quasiperiodic potential ($V_0/E_r = 0.0$), where the only external potential is the harmonic trap. On~the \textbf{right}, the~presence of the quasiperiodic potential is reflected in the distribution of the polymers, which tend to localize at the minima.
\label{fig-1}}
\end{figure}

As motivated in Section~\ref{sec:1}, we aim to discuss the utilization of PIMC methods, implementing zonal estimators for the quantum properties, in~systems displaying a non-periodic patterning. 
We introduce a model of $N$ identical bosons in continuous two-dimensional space described by the Hamiltonian
\begin{equation}\label{eq:ham}
\mathcal{H} = \sum_{i=1}^{N} \left( \frac{\hat{\textbf{p}}_i^2}{2m} + \frac{m\omega^2}{2} \hat{\textbf{r}}_i^2 + V_{qc}(\hat{\textbf{r}}_i) \right) + \sum_{i<j} V_{int}(|\hat{\textbf{r}}_i-\hat{\textbf{r}}_j|),
\end{equation}
where $m$ is the particle mass, $\hat{\textbf{p}}_i$ and $\hat{\textbf{r}}_i$ are the momentum and position operators of the $i$-th particle, $\omega$ is the frequency of the two-dimensional harmonic trap confining the bosons. $V_{qc}$ is an external potential defined by
\begin{equation}\label{eq:vqc}
V_{qc}(\textbf{r}) = V_0 \sum_{i=1}^{4} \cos^2 (\textbf{k}_i \cdot \textbf{r}),
\end{equation}
\begin{equation}\label{eq:kvectors} 
\textbf{k}_1= k_{lat} \begin{pmatrix} 1 \\ 0 \end{pmatrix} \, , \qquad
\textbf{k}_2= k_{lat}/\sqrt{2} \begin{pmatrix} 1 \\ 1 \end{pmatrix} \, , \qquad
\textbf{k}_3= k_{lat} \begin{pmatrix} 0 \\  1 \end{pmatrix} \, , \qquad
\textbf{k}_4= k_{lat}/\sqrt{2} \begin{pmatrix} 1 \\ -1 \end{pmatrix} \, ,
\end{equation}
with $k_{lat}$ setting the spatial modulation, and~$V_0$ the strength of the external potential. The~structure of maxima and minima of this potential displays the geometry of the aperiodic, eightfold-symmetrical Ammann--Beenker tiling~\cite{gru86}, therefore underlying a quasicrystalline structure.  $V_{int}$ is a contact potential with scattering length $a$. Figure~\ref{fig-1} depicts the $snapshot$  configurations of interacting bosons described by Hamiltonian \eqref{eq:ham} with only the harmonic trap (a) and in the presence of the quasicrystalline potential (b).   

\begin{figure}[t]
\includegraphics[width=10.cm]{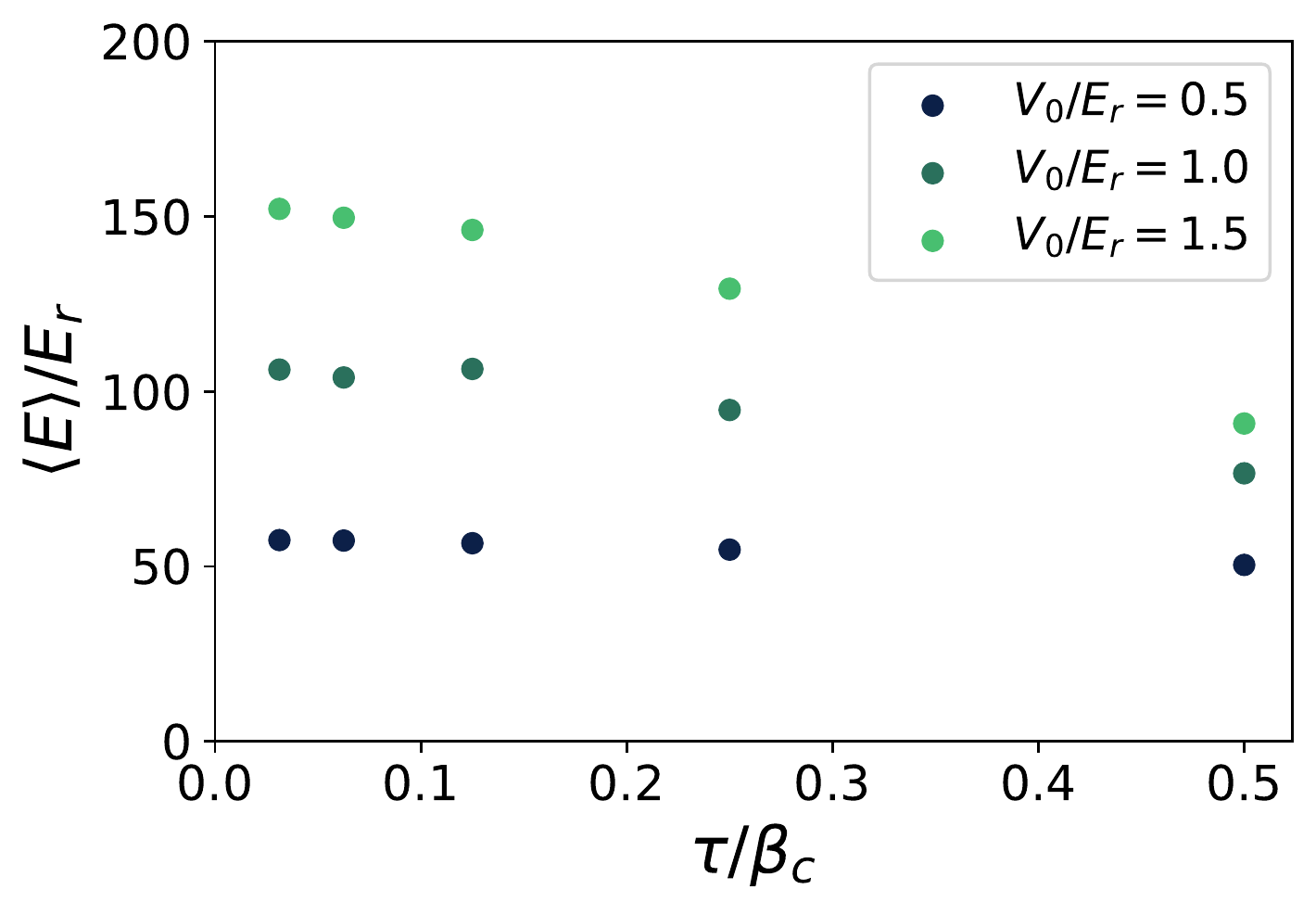}
\caption{Trotter limit. The~circles correspond to the measured values of the average energy $\left\langle E \right\rangle$, at~decreasing values of the imaginary time step $\tau$. Different colors represent different values of the potential: $V_0/E_r = 0.5$ (dark blue), $V_0/E_r = 1.5$ (grey blue), $V_0/E_r = 2.5$ (light blue).
\label{fig1}}
\end{figure}

In Figure~\ref{fig1}, we show the convergence of the thermodynamic average of the total energy as we increase the number of slices $M$, and~reduce the time step $\tau$. Varying the strength $V_0$ at $T=0.25\,T_c$, convergence is already established around $\tau \approx 0.1\beta_c$. Other observables, such as the superfluid fraction, are found to converge even~faster.

We now briefly review the ground-state mean-field approach, which we use to plot density profiles and display the validity of the pair-product approximation. The~system is described by a two-dimensional time-independent Gross--Pitaevskii equation~\cite{pit16},
\begin{equation} \label{eq:gp2d}
    \mu \Psi(\textbf{r}) = \left( -\frac{\hbar^2 \nabla^2}{2m} + \frac{m \omega^2 r^2}{2} + V_{qc}(\textbf{r}) + g |\Psi(\textbf{r})|^2 \right) \Psi(\textbf{r}) \, .
\end{equation}

The two-dimensional mean-field parameter $g$ satisfies, up~to logarithmic corrections,
\begin{equation} \label{eq:g}
g = \frac{\hbar^2}{m} \frac{4\pi}{\ln \left( 4.376 \, / a^2 n(0) \right)} \, ,
\end{equation}
with $n(0)$ the particle density at the center of the two-dimensional trap, and~$a$ the scattering length of the repulsive interaction~\cite{pil05, pet00, pet01}. Contrary to the three-dimensional case, where the relationship between mean-field constant and scattering length is linear, the~logarithm in \eqref{eq:g} implies that $a$ must be vanishingly small for small but finite values of $g$. 

We introduce the harmonic oscillator length $l_{osc} = \sqrt{\hbar / m \omega}$, and~rewrite \eqref{eq:g} as
\begin{equation} \label{eq:g2}
g = \frac{\hbar^2}{m} \tilde{g} = \frac{\hbar^2}{m} \frac{4\pi}{\ln \left( l_{osc}^2 / a^2 \right) + \ln \left( 4.376 \, /  n(0) l_{osc}^2 \right)} \, \approx \, \frac{\hbar^2}{m} \frac{2\pi}{\ln \left( l_{osc} / a \right) } \, ,
\end{equation}
where the last approximate equality applies in the limit $ a/l_{osc} \ll 1$. We use this approximate form to describe the interaction~\cite{gau21}, defining
\begin{equation}
g_0 = \frac{\hbar^2}{m} \tilde{g}_0 = 2\pi \left( \ln  \frac{l_{osc}}{a} \right) ^{-1}.
\end{equation}

In Figure~\ref{fig0}, we plot the density profiles obtained for interacting bosons in a two-dimensional harmonic trap, sliced across the center. The~measured profiles (solid lines) are compared with those predicted by a two-dimensional Thomas--Fermi approximation
\begin{equation}
n(\textbf{r}) = 
\begin{cases}
    \left( \mu - V(\textbf{r}) \right) / g_0 \, , \qquad & V(\textbf{r}) < \mu \\
    0 \, , \qquad & V(\textbf{r}) > \mu
  \end{cases}
\end{equation}

In practice, since in our simulations we work at fixed particle number $N=500$, we first derive $\mu$ as $ \mu = \hbar\omega \sqrt{N \tilde{g}_0 / \pi }. $
We find good agreement between the numerical profiles and analytical ones, even for large values of $g_0$, supporting the validity of the approach used to treat the propagator \eqref{eq:final_rho}. In~the following, we express lengths in units of $l_{osc}$, and~energies in units of $E_r = \hbar^2 k_{lat}^2 / 2m$; we employ a weak harmonic trapping with $\hbar \omega \approx 0.02 \, E_r$. The~connection between the values relative to the harmonic trap and those pertaining to the lattice is the ratio $l_{osc} / \lambda_{lat} \approx 1.6$.

\section{Density Profiles and Diffraction~Patterns}\label{sec:4}

\textls[-19]{In the present section, we introduce the physical estimators obtained via PIMC, and~we discuss the results achieved for trapped bosons in a quasiperiodic potential in two~dimensions. }

There are several complementary ways to display spatial configuration of the quantum system and its classical-polymer equivalent system. One is to select a system's configuration at a given simulation step, plotting the position of the beads $\textbf{r}_{i,m}$: the resulting snapshots provide a first graphical estimate of the particle's probability distribution. 
In Figure~\ref{fig-1} we show snapshots of particle configurations at $\tilde{g}_0=2.1704$, for~two values of $V_0$. Red lines represent the links between successive beads in the equivalent polymer system, as~described in Section~\ref{sec:3}. The~shaded background represents the potential, with~darker to brighter areas corresponding to lower to higher values. On~the left, we show a configuration in the absence of the quasiperiodic potential ($V_0/E_r = 0.0$), where the only external potential is the harmonic trap. On~the right, the~presence of the quasiperiodic potential is reflected in the distribution of the polymers, which tend to localize at the~minima.

\begin{figure}[t]
\includegraphics[width=0.9\linewidth]{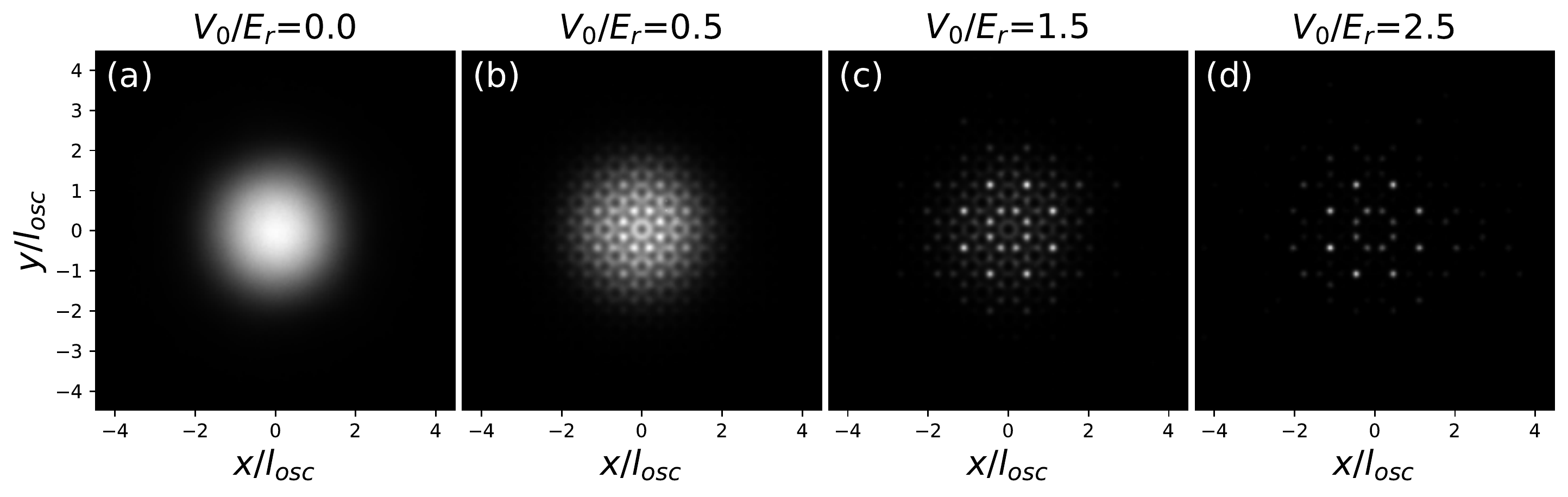}
\caption{{Density} patterns at $\tilde{g}_0 = 0.0217$ and $T/T_c = 0.25$. In~each picture, the~normalized two-dimensional density takes on values going from 0 (black) to 1 (white). The~density patterns are shown at four values of $V_0$: $V_0/E_r=0.0$ (\textbf{a}), $V_0/E_r=0.5$ (\textbf{b}), $V_0/E_r=1.5$ (\textbf{c}), and $V_0/E_r=2.5$ (\textbf{d}).
\label{fig3}}
\end{figure}  

Our approach give access to density profiles, which are obtained as averages over simulation steps, as~well as over the positions of all different beads associated to each particle. In~continuous space, the~average is performed  by separating the simulation area into bins, and~counting the number of beads in each at every simulation step. 
In Figure~\ref{fig3}, we show two-dimensional density patterns for the system at $\tilde{g}_0 = 0.0217$ and $T/T_c = 0.25$. Figure~\ref{fig3}a is the density profile in the harmonic trap, the~same shown in Figure~\ref{fig1}. As~the value of $V_0$ increases, the~quasicrystalline structure appears initially as a modulation (b) and then as localization at the deepest minima (c--d). 

Diffraction patterns can be investigated employing the structure factor that, for~example, is observed experimentally in scattering experiments. For~a particle density distribution
\begin{equation}\label{distn}
n(\textbf{r}) = \sum_i \delta(\textbf{r}_i)\,, 
\end{equation}
the structure factor reads
\begin{equation}
	\label{struct}
	I(\textbf{q}) = \langle n(\textbf{q}) n(-\textbf{q}) \rangle, 
\end{equation}
where  $n(\textbf{q})$ is defined as following
\begin{equation} 
n(\textbf{q}) = \int d^2\textbf{r} \; e^{-i\textbf{q}\cdot\textbf{r}} n(\textbf{r}) = \sum_{j} e^{-i\textbf{q}\cdot\textbf{r}_j} 
\end{equation}
which is the Fourier transform of the particle distribution \eqref{distn}~\cite{cha95}. 

\begin{figure}[t]
\includegraphics[width=0.9\linewidth]{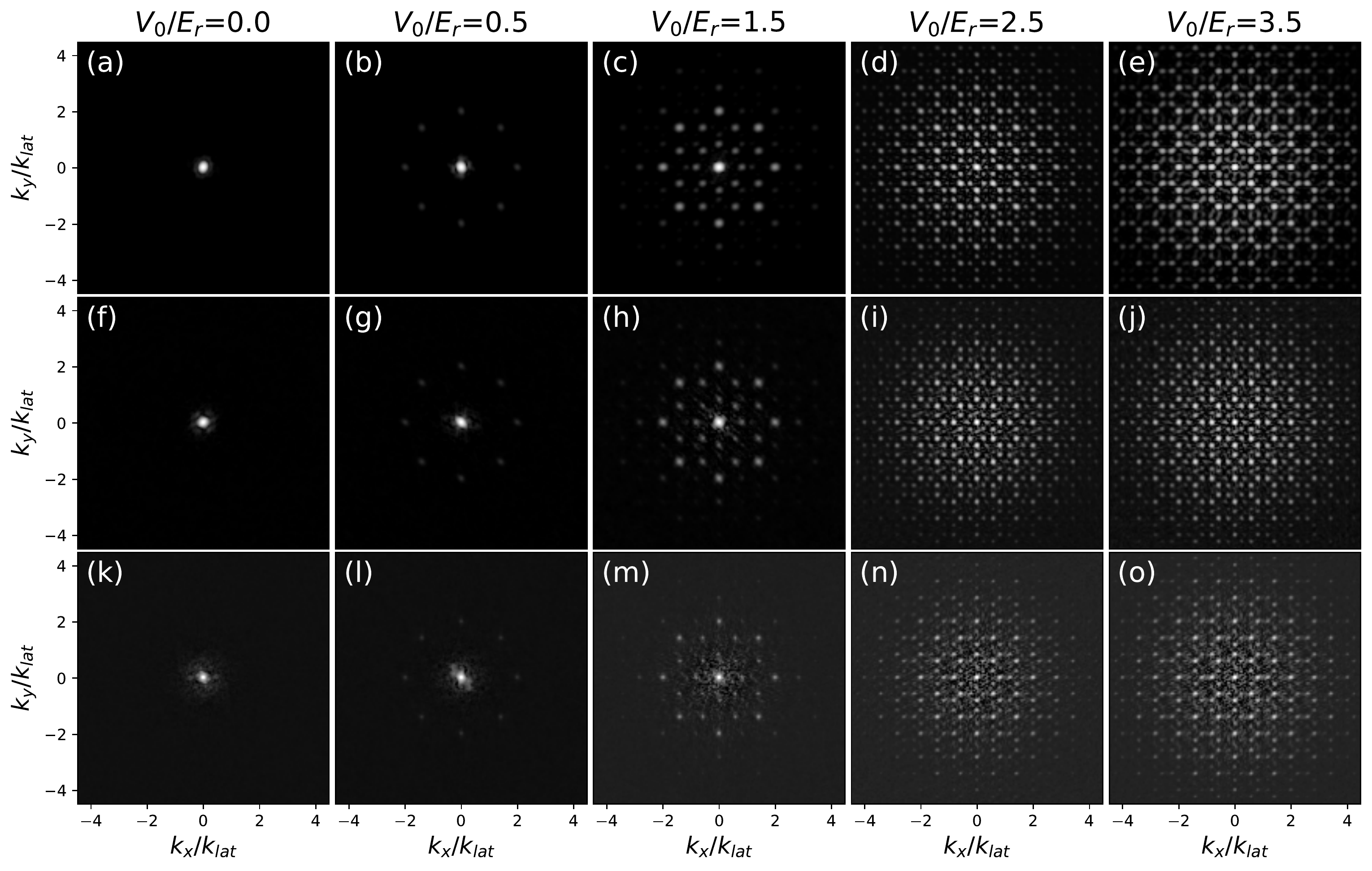}
\caption{Diffraction patterns at $\tilde{g}_0 = 0.0217$. In~each picture, the~density in $k$-space is shown, taking on values from a lower cutoff (black) to the maximum (white). Each column displays a value of $V_0$: $V_0/E_r=0.0$ (\textbf{a},\textbf{f},\textbf{k}), $V_0/E_r=0.5$ (\textbf{b},\textbf{g},\textbf{l}), $V_0/E_r=1.5$ (\textbf{c},\textbf{h},\textbf{m}), $V_0/E_r=2.5$ (\textbf{d},\textbf{i},\textbf{n}), $V_0/E_r=2.5$ (\textbf{e},\textbf{j},\textbf{o}). Each row displays a value of $T$: $T/T_c=0.25$ (\textbf{a}--\textbf{e}), $T/T_c=0.4$ (\textbf{f}--\textbf{j}), $T/T_c=0.7$ (\textbf{k}--\textbf{o}).\label{fig4}}
\end{figure}

Figure~\ref{fig4} displays some examples of diffraction patterns. As~we should expect, the~structure factor evolves from a single peak in the fluid phase to a typical quasicrystalline pattern as $V_0$ increases. The~three rows in the figure correspond to three different temperatures; we can see that, aside for some smearing of the peaks due to thermal fluctuations, the~structure remains essentially unchanged. This is expected since, even above $T_c$, strong intensities of the lattice lead to localization, and~the formation of a classical~insulator.


\section{Global Quantum~Estimators}\label{sec:5}

\begin{figure}[t]
\includegraphics[width=10.cm]{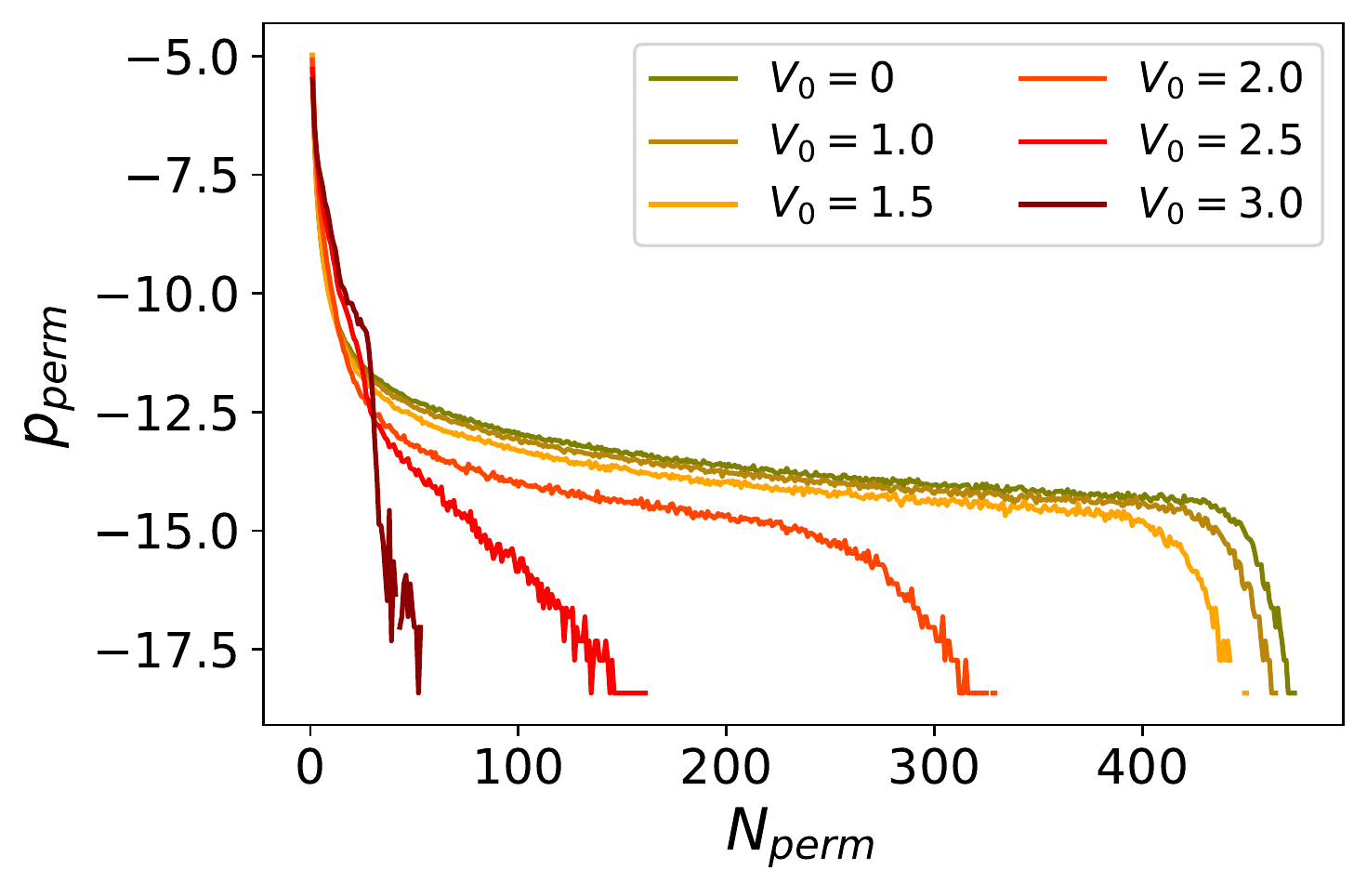}
\caption{Probabilities of permutation cycles $P_{perm}$ as a function of cycle length, at~$\tilde{g}_0 = 0.0217$ and $T/T_c = 0.25$. Each color corresponds to a different value of $V_0$, as~indicated in the legend. Although~histograms are generally more appropriate, we use solid lines for ease of comparison between different cases. All plots are in logarithmic~scale. \label{fig6}}

\end{figure}

A first impression of the importance of quantum effects in the simulation can be obtained by looking at particle permutations~\cite{PhysRevB.84.014534}. As~mentioned above, due to the bosonic nature of the particles, the~polymers of the equivalent classical system connect to each other, forming long cycles containing varying number of particles. We call the number of particles in a cycle $N_{perm}$. At~each simulation step, we can construct a histogram, counting how many cycles contain exactly $N_{perm}$ particles for each value of $0 < N_{perm} \le N$. We can then average the histogram over simulation steps, and~normalize it, so that for each value of $N_{perm}$ we obtain the probability $p_{perm}$ of finding a cycle with exactly $N_{perm}$ particles.  The~histogram of $p_{perm}$ is shown in Figure~\ref{fig6} for different values of $V_0$. The~distribution of $p_{perm}$ in the superfluid phase shows that permutations entail cycles comprising almost all particles in the trap. 
Intermediate values of $V_0$ lead to the disappearance of permutation cycles comprising of all particles in the system, but~still allow for cycles of a few hundred particles. Even larger values lead to a sharp drop in $p_{perm}$ as a function of $N_{perm}$, with~only cycles of the order of ten particles remaining~relevant. 

We further characterize the quantum regime by considering the superfluid fraction. We recall that, in~the context of the two-fluid model~\cite{tis38, lon54,cha95}, the~density of a quantum system displaying superfluidity at low temperature can be described by decomposing its density $\rho$ into a sum of two fields: $\rho = \rho_n + \rho_s$, where the first component describes the normal density, whereas the second is the superfluid one. In~this context, the~superfluid fraction is defined as the ratio of the superfluid density to $\rho$:
\begin{equation} \label{ns0}
n_s = \frac{\rho_s}{\rho} \,.
\end{equation}

The two components exhibit contrasting behavior in terms of, for~example, flow transport and entropy~\cite{Leggett2006}. The~superfluid component displays zero viscosity, and~is therefore unresponsive to the application of external velocity fields. In~particular, when subjected to an angular velocity field, the~superfluid exhibits a reduction of the total moment of inertia compared to a classical fluid in the same conditions. This phenomenon is encoded in a fundamental relation, which links $n_s$ to the system's moments of inertia:
\begin{equation}\label{ns1}
n_s = 1 - \frac{I}{I_{cl}}\,.
\end{equation}

\textls[-10]{Here, we indicated $I$ as the moment of inertia related to $\rho_n$, $I_{cl}$ representing the $classical$ moment of inertia, which is the one the same mass of fluid would have if it behaved~classically.}

By applying linear response theory on Equation~\eqref{ns1}, one can extract $n_s$ through PIMC. In~fact, it is possible to show that the expectation value of the angular momentum in the quantum system is given in terms of the area encircled by tangled paths in the classical system of polymers~\cite{szi89,zen14}. This way to evaluate Equation~\eqref{ns1} results as particularly appropriate for all trapped and finite-size bosonic systems. Since we are dealing with a pure two-dimensional system, we are interested in studying infinitesimal rotations around an axis perpendicular to the $xy$ plane. Following the formulation given in Ref.~\cite{szi89}, $n_s$ in its complete form reads
\begin{equation} \label{global}
n_s = \frac{4 m^2}{\hbar^2 \beta I_{cl} }\; \left(\langle A_z^2 \rangle - \langle A_z \rangle^2\right)\,,
\end{equation}
where $A_z$ is the component of the total area enclosed by particle paths on the $xy$ plane, defined as
\begin{equation}
A_z =  \frac{1}{2} \sum_{i=1}^{N} \sum_{m=0}^{M-1} \left(\textbf{r}_{i}^{m} \times \textbf{r}_{i}^{m+1}\right)_z\,,
\end{equation}
with $\textbf{r}_i^m$, the position of the bead corresponding to the $i$-th particle on the $m$-th time slice, the~same as in Section~\ref{sec:3}, whereas the classical moment of inertia in Equation~\eqref{global} is computed as
\begin{equation}
I_{cl}= m\; \bigg \langle  \sum_{i=1}^{N} \sum_{m=0}^{M-1} \textbf{r}_{i}^{m} \cdot \textbf{r}_{i}^{m+1}   \bigg \rangle\,.
\end{equation}

Usually, the $\langle A_z \rangle$ term is set to zero, or~ignored altogether~\cite{Ceperley1995}. The~most convincing argument is that, from~an energetic point-of view, and~therefore as far as equilibrium probabilities are concerned, any configuration is equivalent to a symmetric one where the directions of all links between nearest-neighbor beads have been reversed; if both configurations can be accessed, they should be visited with equal probability. Since reversing all links also changes the sign in the area corresponding to the configuration, the~immediate consequence is that the average value of the area will be zero, leading to $\langle A_z \rangle=0$. However, in~configurations where bosons are localized into clusters (such as the deepest minima of the quasiperiodic optical lattices, e.g.,~see Figure~\ref{fig-1}), this term does not necessarily average to zero, and~it must be kept into account. This observation is justified by the fact that the system might spend a long period of time (compared to simulation times) in a region of configuration space where $\langle A_z \rangle\neq0$, leading to a manifestation of ergodicity~breaking.

\begin{figure}[t]
\includegraphics[width=10.cm]{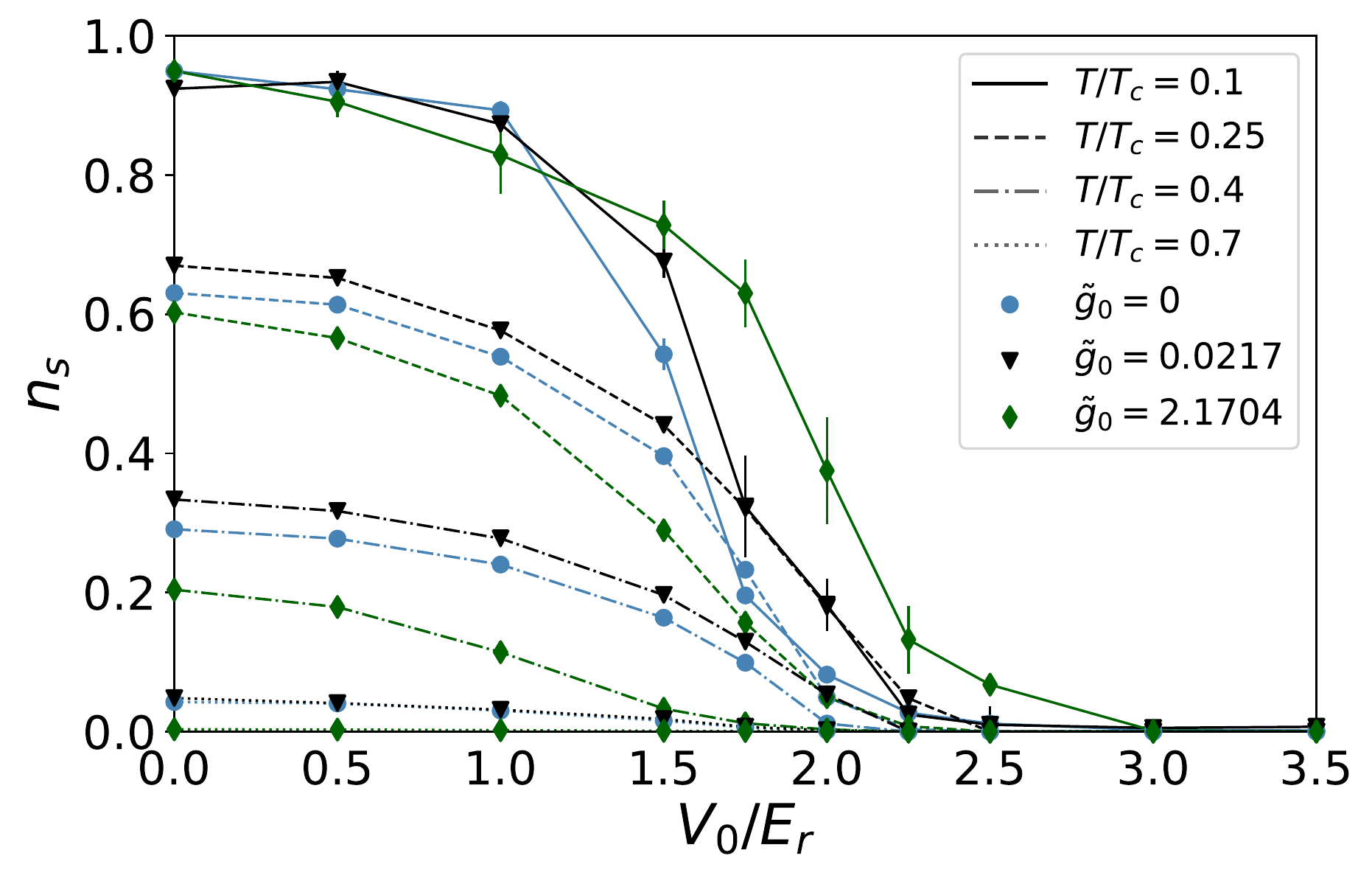}
\caption{Measured values of the global superfluid fraction at different values of $V_0$. Colored markers correspond to different values of the interaction: $\tilde{g}_0 = 0$ (teal circles), $\tilde{g} = 0.0217$ (black triangles), $\tilde{g}_0 = 2.1704$ (green diamonds). Lines are a guide for the eye, and~also serve to distinguish different temperatures: $T/T_c=0.1$ (solid lines), $T/T_c=0.25$ (dashes), $T/T_c=0.4$ (dashes and dots), $T/T_c=0.7$ (dots). 
\label{fig7}}
\end{figure}

In Figure~\ref{fig7}, we show plots of $n_s$ as a function of $V_0$, at~different values of temperature and  interaction strength. In~all cases, deeper quasiperiodic lattices bring about a reduction of the global superfluidity, up~to a critical value at which it is completely depleted, and~the system transitions to a localized~phase.

\section{Zonal Superfluid~Estimators}\label{sec:5a}

In dealing with inhomogeneous systems, it also worthwhile to extrapolate superfluid features that may be spatially dependent, introducing density fields $\rho(\textbf{r})$, $\rho_s(\textbf{r})$, and~$\rho_n(\textbf{r})$, which extend the uniform quantities introduced above. We can then combine Equation~\eqref{ns0} and Equation~\eqref{ns1}, and~the definition of the classical moment of inertia $I_{cl} = \int d\textbf{r} \; \rho(\textbf{r}) r^2$ to~write
\begin{equation} \label{nslocal}
n_s = \frac{1}{I_{cl}} \int d\textbf{r} \; \rho_s(\textbf{r}) r^2 \, , 
\end{equation}
with $r$ representing the distance from the center of the coordinate system. The~definition of a local superfluid fraction follows from \eqref{ns0}, as~$n_s(\textbf{r}) = \rho_s(\textbf{r})/\rho(\textbf{r})$; with this, we can rewrite \eqref{nslocal} as
\begin{equation} \label{nsdecomp}
\frac{1}{I_{cl}} \int d\textbf{r} \; n_s(\textbf{r}) \rho(\textbf{r}) r^2\,,
\end{equation}
meaning that the global superfluid fraction is an average of the local superfluid fraction, weighted by the local moment of inertia. While the introduction of inhomogeneous density fields is natural, their expression in terms of of the classical polymers requires some elaboration; following Kwon~et~al.~\cite{PhysRevB.74.174522}, who have introduced a physically motivated and consistent definition of $\rho_s(\textbf{r})$, we write
\begin{equation}\label{rhoslocal}
	\rho_s(\textbf{r}) = \frac{4 m^2}{\hbar^2 \beta}\frac{ \langle A_z A_z(\textbf{r}) \rangle - \langle A_z \rangle \langle A_z(\textbf{r}) \rangle }{r^2} \, ,
\end{equation}
where
\begin{equation}\label{azlocal}
	A_z(\textbf{r}) =  \frac{1}{2} \sum_{i=1}^{N} \sum_{m=0}^{M-1} \textbf{r} \times \textbf{r}_{i}^{m+1} \delta(\textbf{r}-\textbf{r}_{i}^{m}). 
\end{equation}

Equations~\eqref{nslocal}, \eqref{rhoslocal} and \eqref{azlocal} allow us to investigate the superfluid behavior locally, usually by sampling \eqref{rhoslocal} on a grid. However, and~especially for strongly inhomogeneous systems such as \eqref{eq:ham}, the~amount of detail proves to be excessive, and~the estimators too noisy. In~order to overcome this issue, and~to extract some degree of spatial information about the system, we act on a middle level by introducing a zonal~estimator.

We divide the system into $K$ regions, labeled by $k=1,2\ldots K$, and~introduce a zonal superfluid fraction by specializing \eqref{nslocal} to
\begin{equation}
n_{s,k} = \frac{1}{I_{cl,k}} \int_k d\textbf{r} \; \rho_s(\textbf{r}) r^2 \, ,
\end{equation}
where $I_{cl,k}$ is the fraction of the total classical moment of inertia corresponding to region $k$, so that $I_{cl} = \sum_{k=1}^K I_{cl,k}$. The~zonal superfluid fractions can be recombined, using \eqref{nsdecomp}, to~give
\begin{equation}\label{decomp}
n_s = \sum_{k=1}^{K} \frac{I_{cl,k}}{I_{cl}} n_{s,k}.
\end{equation}

\begin{figure}[t]
\begin{center}
\includegraphics[width=14.cm]{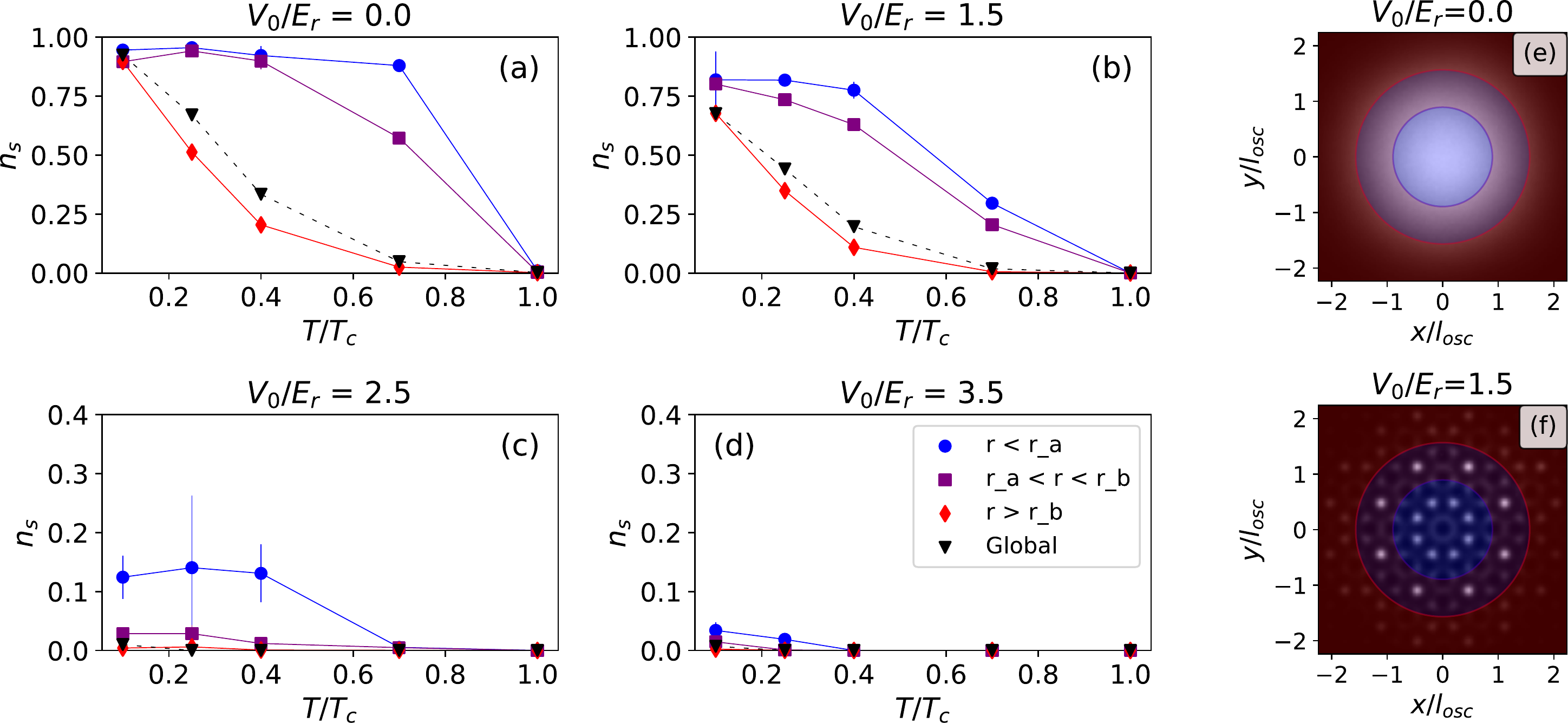}
\caption{Temperature behavior of the superfluid fraction in different regions, at~$\tilde{g}_0 = 0.0217$. The~regions are depicted on the right, where they are superimposed to the density profiles of Figure~\ref{fig3}, for~$V_0/E_r=0.0$ (\textbf{e}) and $V_0/E_r=1.5$ (\textbf{f}). The~inner radius is $r_a$ (purple line) and the outer radius is $r_b$ (red line). Colored markers in the plots (\textbf{a}--\textbf{d}) correspond to values of $n_{s,k}$ measured in different regions: $r < r_a$ (blue circles), $r_a < r < r_b$ (purple squares), $r > r_b$ (red diamonds). The~same colors are used to shade the regions in (\textbf{e},\textbf{f}). We also report the global superfluid fraction, which was already displayed in Figure~\ref{fig7} (black triangles). The~four plots each correspond to a different value of $V_0$: $V_0/E_r = 0.0$ (\textbf{top left}), $V_0/E_r = 1.5$ (\textbf{top right}), $V_0/E_r = 2.5$ (\textbf{bottom left}), $V_0/E_r = 3.5$ (\textbf{bottom~right}).
\label{fig8}}
\end{center}
\end{figure}

In the present case, we exploit the circular symmetry provided by the trap to divide the system into three concentric belts.
It is important to stress that, for~instance, the~decomposition \eqref{decomp} may display  sectors with a finite superfluid fraction, but~still give a negligible contribution to the global $n_s$, if~the associated moment of inertia is small. This is the case for regions close to the trap center, as~we show in Figure~\ref{fig8}, which displays the superfluid fraction $n_{s,k}$ in different regions against~temperature.

Most familiar is the behavior in the case of $V_0$, in~Figure~\ref{fig8}a, where we can see the global superfluidity drop from nearly $n_s=1$ at low temperature to $n_s=0$ at the critical temperature $T_c$. The~depletion of superfluidity proceeds at different rates across the trap: the inner regions display a value of $n_{s,k}$, which is still close to 1, even at high values of $T$, such as $T/T_c=0.7$. The~global superfluidity, however, is dominated by the contribution of bosons in the outer region, due to their larger classical moment of~inertia.

As $V_0$ increases, the~effects of the quasiperiodic lattice become prominent; the global superfluidity is depleted by localization, as~already shown in Figure~\ref{fig7}, but, for~some values of $V_0$, the~zonal superfluidity remains finite in the central region. In~\cite{cia22}, we used this information, coupled with the fluctuations in particle number, to~characterize a Bose Glass phase induced by the quasiperiodic~potential. 

\section{Discussion and~Conclusions} \label{sec:6}

The present work we detailed PIMC methods to explore both local and global quantum properties of interacting bosons confined in external quasiperiodic potential. Those properties have been probed in a finite temperature regime with specific attention to superfluidity. In~detail, 
we summarized the derivation of the pair-product approximation for particles interacting through hard-core interactions in two or three dimensions, and~we presented the details of our implementation. Then, we explored a new zonal estimator, which gives access to local information about the superfluidity in finite regions of trapped systems, and~it is therefore well-suited to the study of spatially inhomogeneous potentials. 
For the example presented in this work, zonal estimators are relevant to the detection of correlated phases, such as the Bose glass phase, which is characterized by rare regions where superfluidity and finite compressibility coexist~\cite{cia22}.
Similar zonal estimators can be applied to other quantities, such as regional fluctuations of particle number, associated with density compressibility, and~magnetic ordering in multi-component systems or spin systems related to the spin~compressibility. 

Moreover, one might apply such estimators to the characterization of local properties of self-assembled quasicrystalline phases in free space generated by two-body nonlocal interactions~\cite{defenu2021longrange}.
A prime example of a two-body model potential leading to quasiperiodic patterns is in the paradigmatic hard-soft corona potential, which is largely used to investigate purely classical systems~\cite{Malescio:2003aa}.
The same model has also been applied to bosonic systems, where the effects of zero point motion, as~well as quantum exchanges, disclose rich phase diagrams including quantum quasicrystal with $12$-fold rotational symmetry~\cite{Barkan2014}. Additionally,
quantum properties of self-assembled cluster quasicrystals revealed that, in~some cases, quantum fluctuations do not jeopardize dodecagonal structures, showing a small but finite local superfluidity~\cite{PhysRevB.101.134522}. 
Cluster quasicrystals display peculiar features, not exhibited by simpler quasiperiodic structures. By~increasing quantum fluctuations, in~fact, a~structural transition from quasicrystal to cluster triangular crystal featuring the properties of a supersolid is observed~\cite{PhysRevB.101.134522,Fa2019,Cinti2019}. We point out that the discussed methodology is also useful to analyze the superfluid character of further peculiar inhomogeneous systems such as, for~instance, bosons enclosed within spherical traps or subject to a polyhedral-symmetric substrate potential~\cite{PhysRevLett.125.010402,PhysRevA.103.033313,app112110053}.

\authorcontributions{Conceptualization, Matteo Ciardi, Tommaso Macrì and Fabio Cinti; Data curation, Matteo Ciardi, Tommaso Macrì and Fabio Cinti; Formal analysis, Matteo Ciardi, Tommaso Macrì and Fabio Cinti; Investigation, Matteo Ciardi, Tommaso Macrì and Fabio Cinti; Methodology, Matteo Ciardi, Tommaso Macrì and Fabio Cinti; Writing – original draft, Matteo Ciardi, Tommaso Macrì and Fabio Cinti. All authors have read and agreed to the published version of the manuscript.}

\funding{T.M. acknowledges CNPq for support through Bolsa de produtividade em Pesquisa n.311079/2015-6 and support from CAPES. This work was supported by the Serrapilheira Institute (grant number Serra-1812-27802).}

\institutionalreview{Not applicable.}
%
\informedconsent{Not applicable.}

\dataavailability{Not applicable.}

\acknowledgments{We acknowledge V. Zampronio for discussions and for a careful reading of the manuscript. T. M. acknowledges the hospitality of ITAMP-Harvard where part of this work was done. We thank the High Performance Computing Center (NPAD) at UFRN and the Center for High Performance Computing (CHPC) in Cape Town for providing computational resources.}

\conflictsofinterest{The authors declare no conflict of interest.}

\begin{adjustwidth}{-\extralength}{0cm}

\reftitle{References}


\end{adjustwidth}

\end{document}